# Probing the magnetism of topological end-states in 5-armchair graphene nanoribbons

James Lawrence,[1,2,†] Pedro Brandimarte,[1,†] Alejandro Berdonces,[1,2] Mohammed S. G. Mohammed,[1,2] Abhishek Grewal,[3] Christopher C. Leon,[3] Daniel Sanchez-Portal,[1,2,*] Dimas G. de Oteyza[1,2,4,*]

[1] Donostia International Physics Center, 20018 San Sebastián, Spain

[2] Centro de Fisica de Materiales, CSIC-UPV/EHU, 20018 San Sebastián, Spain

[3] Max-Planck-Institut für Festkörperforschung, 70569 Stuttgart, Germany

[4] Ikerbasque, Basque Foundation for Science, 48011 Bilbao, Spain

[†] These authors contributed equally

[*] daniel.sanchez@ehu.eus, d_g_oteyza@ehu.eus

**Abstract:**

We extensively characterize the electronic structure of ultra-narrow graphene nanoribbons (GNRs) with armchair edges and zig-zag termini that have 5 carbon atoms across their width (5-AGNRs), as synthesised on Au(111). Scanning tunnelling spectroscopy measurements on the ribbons, recorded on both the metallic substrate and a decoupling NaCl layer, show well-defined dispersive bands and in-gap states. In combination with theoretical calculations, we show how these in-gap states are topological in nature and localised at the zig-zag termini of the nanoribbons. Besides rationalising the driving force behind the topological class selection of 5-AGNRs, we also uncover the length-dependent behaviour of these end states which transition from singly occupied spin-split states to a closed-shell form as the ribbons become shorter. Finally, we demonstrate the magnetic character of the end states via transport experiments in a model two-terminal device structure in which the ribbons are suspended between the scanning probe and the substrate that both act as leads.

**TOC:**

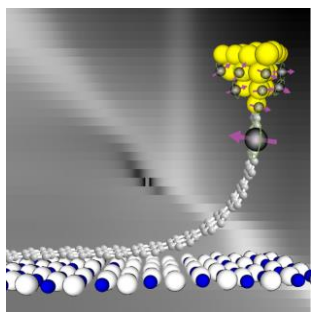

The on-surface synthesis and characterisation of graphene nanoribbons (GNRs) has been a source of intense research interest in recent years[1–4]. Extensive efforts have been made to understand the relationship between their width/edge structure and their electronic properties[5–7]. Many of these studies involve the utilisation of metal-catalysed on-surface synthetic methods, as well as characterisation via typical surface science methods, including scanning tunnelling microscopy and spectroscopy (STM/STS). GNRs with so-called 'armchair' edges (AGNRs) are particularly well-studied, with experimental band gap values reported for 5-[8,9], 6-[5], 7-[10–12], 8-[13], 9-[5,14], 10-[13], 12-[5] and 13-AGNRs[12] (with the prefix equal to the number of carbon atoms across the width of the ribbon).

When comparing these values to those predicted by theory, there is good agreement when the GW formalism is used with an added semi-empirical substrate screening.[6] However, there are still notable exceptions. The measurements on 5-AGNRs in particular show a very large disagreement with theory, with one study estimating the measured band gap at 2.8 eV,[9] and another giving a substantially lower value of 0.1 eV.[8] In contrast to this, soluble 5-AGNRs (that are functionalized with polar side groups) have been shown to have an optical band gap of 0.88 eV[15]. Other articles have also made the claim that (3p+2)-GNRs are metallic[16,17], as supported by tight-binding calculations. However, these studies failed to note the effects of the relaxation of the ribbon structure[18] and the potential for topological effects at the boundaries of these ribbons – in particular, the possibility of forming edge (or 'end') states that may be observed in spectroscopic experiments, substantially complicating their interpretation.

Topological phenomena in graphene nanoribbons have only recently become a subject of interest in the field. In particular, topological states that exist due to the formation of boundaries such as intra-ribbon junctions[19,20] and the finite termination of nanoribbons[21,22] have been observed, with presumed spin-split states often associated with zig-zag edge topologies[21,23–25]. Engineering topological states through the controlled growth of nanostructures also enables the formation of topological bands[19,20], allowing an increased level of precise control over the electronic structure of such materials. The existence of these states may be predicted by computing their Zak phase, with a characteristic $\mathbb{Z}_2$ value of 0 or 1 for topologically trivial and non-trivial phases, respectively. A more detailed explanation, along with several salient examples, may be found in the article of Cao et al.[7]

In the following, we have further investigated the nature of 5-AGNRs via scanning tunnelling spectroscopy/microscopy and density functional theory. In particular, we have examined the electronic structure of ribbons of different lengths adsorbed directly on Au(111), as well as on decoupling NaCl monolayer islands. The subtle role of the anisotropic electrostatic potential felt by valence electrons along the sides of the GNR is also discussed and provides an explanation for the topological class selection of 5-AGNRs. This leads to the appearance of in-gap end states at the ribbon termini and explains the apparently narrow band gap observed in previous measurements[8]. To provide additional proof for this interpretation, we have undertaken nanoribbon lifting experiments in which the conductance of the ribbons is measured as a function of lifting height via contact with an STM tip. The appearance of a Kondo resonance in this model two-terminal device geometry proves the magnetic nature of the topological end states. Finally, we have also demonstrated that simply exposing these ribbons to molecular oxygen results in the destruction of the end states. Half-oxidized ribbons with only one end state have allowed us to determine the origin of the measured spectroscopic resonances.

**Results and Discussion:**

In order to investigate the properties of the 5-AGNRs, they were first synthesised on Au(111) by depositing a low coverage of the precursor molecule, dibromoperylene (DBP, structure shown in Fig.

1(a)), onto the substrate whilst it was held at 418 K (see methods). The resulting ribbons are shown in Fig. 1(b). Both straight and kinked ribbons (via an alternate fusing mechanism, as discussed by Kimouche et al.[8]) are observed; however the focus of this study is purely on the regular straight 5-AGNRs. NaCl and CO were post-deposited to provide decoupling layers and tip functionalization options, respectively.

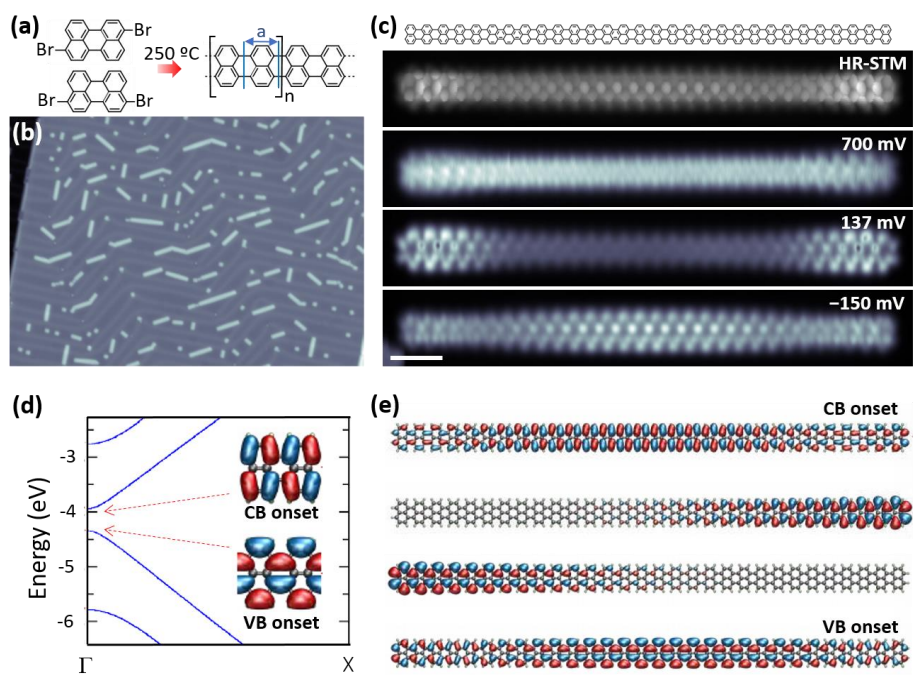

**Figure 1.** (a) Molecular structure of the reactants (present in equal amounts) and the resulting end product (5-AGNR) with its associated unit cell (UC) of dimension *a*. (b) Large scale STM image (I = 50 pA, U = -0.5 V) of a representative GNR-covered Au(111) sample. (c) The chemical structure of a 30-UC 5-AGNR, a constant height HR-STM[26] image of the ribbon (5 mV, CO tip) and constant height dI/dV images of three states found near the Fermi level (CO tip, oscillation: f = 731 Hz, A = 15 mV, energy displayed on images). Scale bar 1 nm. (d) DFT-calculated band dispersion of fully relaxed free-standing 5-AGNRs and the wavefunctions of electronic states at valence and conduction band onsets. (e) DFT-calculated gas phase molecular orbitals for a 30-UC long 5-AGNR. The orbitals displayed are the first molecular orbital associated with the valence band (VB), the conduction band (CB), as well as additional spin-polarized in-gap states (only the singly occupied states are displayed, refer to Fig. S1 to see a larger set of orbitals that also includes the singly unoccupied states).

When initially recording STM/STS of a 30-unit cell (UC) long ribbon, there are three states close to the Fermi level that are immediately apparent, as shown in the constant height dI/dV images in Fig. 1(c), along with a high resolution-STM (HR-STM) image that reveals the chemical structure of the ribbon. After examining the DFT calculated band structure of infinite 5-AGNRs (Fig. 1(d)), it becomes apparent (taking into consideration the p-wave nature of our CO-functionalized probe[27]) that the states observed at -150 mV and at 700 mV are associated with the VB and CB onsets, respectively. The state at 137 mV, notably confined to the GNR end regions, therefore corresponds to an in-gap state that is not predicted by calculations performed on infinite ribbons. Instead, calculations on finite 30-UC ribbons (Fig. 1(e)) reproduce not only the VB-related and CB-related states, but also spin-polarized in-gap states whose superposition nicely resembles the experimental data. From the combination of these experimental and theoretical data it is thus immediately clear that the actual band gap of a relatively long 5-AGNR on Au(111) is around 0.85 eV, that the ribbon is not metallic, and that it displays additional in-gap states whose nature will be discussed in the following.

According to standard tight binding models, in which all bonds are characterized by the same hopping constant t, 5-AGNRs (and the whole family of AGNRs with 3p+2 atoms across their width) are predicted to be metallic.[18] This result, however, does not consider the chemically different nature of the C atoms along the edge of the nanoribbons, nor the relaxations brought about by such edge effects. In order to exemplify these effects we model 5-AGNRs by including not one, but two different hopping constants: one for the bonds along the longitudinal ribbon axis ($t_{par}$) and another for the transversal bonds ($t_{perp}$). The results are displayed in Fig. 2(a) and show how, starting from a zero bandgap ribbon at $t_{par}=t_{perp}$ conditions, its bandgap opens as the two hopping constants become increasingly different. A topological analysis of the two semiconducting band structures reveals that they display different $\mathbb{Z}_2$ invariant values, being topological ($\mathbb{Z}_2 = 1$) for $t_{par}>t_{perp}$ and trivial ($\mathbb{Z}_2 = 0$) for $t_{par}<t_{perp}$. This difference is important to understand the ultimate properties of 5-AGNRs because topological ribbons will develop in-gap states at their ends as they are made finite, while trivial ribbons will not. It is worth noting here that this topological behavior is reproduced in the calculations even if only the hopping constants corresponding to parallel bonds between the singly hydrogenated carbon atoms at both sides of the ribbon are made different from the rest.

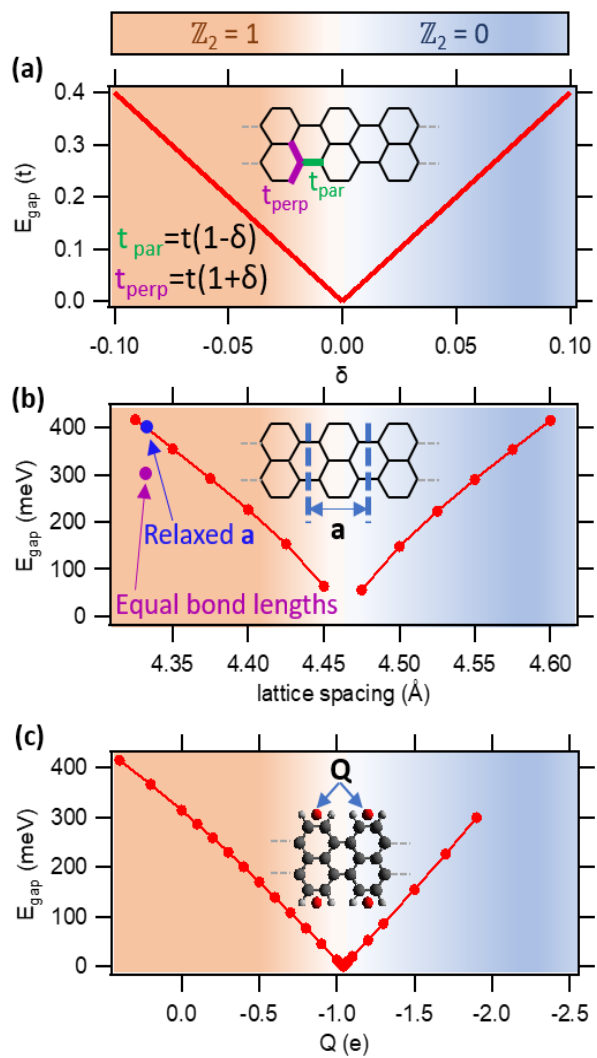

**Figure 2.** Topological class ($\mathbb{Z}_2 = 0$ or $\mathbb{Z}_2 = 1$) and band-gap of 5-AGNRs as a function of different parameters: (a) As a function of the varying hopping constants $t_{par}$ and $t_{perp}$, marked in the inset, in a simple $\pi$ tight-binding model. (b) As a function of the lattice spacing **a** in a DFT calculation. The lowest energy lattice spacing as obtained from DFT is marked with a blue point, while the purple data point marks the band gap value at that same lattice spacing but forcing all C-C bonds to be equally long (C-H bond lengths relaxed). (c) As a function of the point charges Q positioned as marked in the inset (calculated for a = 4.36 Å) at DFT level.

It is well known that it is energetically favourable for materials to open a bandgap, and that scenario is thus also expected for 5-AGNRs, but it is not intuitively clear whether it will open into the regime represented by $t_{par}>t_{perp}$ or by $t_{par}<t_{perp}$. To further investigate this question, we have performed more realistic DFT calculations with varying unit cell lattice spacings (Fig. 2(b)). The results show a similar trend to that observed in Fig. 2(a). There is a topological inversion point with a negligible band gap at a lattice spacing of 4.46 Å and the band gap opens towards the trivial ($\mathbb{Z}_2 = 0$) or topological ($\mathbb{Z}_2 = 1$) class as the lattice spacing increases or decreases from 4.46 Å, respectively. An intuitive analogy can be drawn with respect to the tight binding calculations of Fig. 2(a). Since longitudinal stretching affects the parallel bonds more, shorter or larger lattice spacings will promote a stronger or weaker $t_{par}$ coupling relative to that of the transversal bonds $t_{perp}$, respectively. Fig. S2 compares, for 5-AGNRs with lattice parameters of 4.33 Å and 4.60 Å, the wavefunctions at the $\Gamma$ point and at the Brillouin zone edge for each of the low energy bands. As expected from a topological inversion, the VB and CB at their onset (at $\Gamma$) appear exchanged for either topological class. Instead, at the zone boundary, as well as anywhere along the higher energy bands, the orbitals all appear unchanged. We have also performed similar calculations on finite, 30 UC long 5-AGNRs (Fig. S1), revealing again the exchanged wavefunction symmetry between the occupied and unoccupied orbitals near $E_F$, as well as the presence or absence of end states depending on the topological class.

The lowest energy configuration corresponds to an equilibrium lattice parameter of 4.33 Å (marked in Fig. 2(b) with the blue point), which is clearly on the topological side and displays a notable bandgap. However, a substantial band gap value and the same topological class are obtained from DFT calculations of 5-AGNRs in which the 4.33 Å lattice parameter is maintained, but all bond lengths are forced to be equal (purple point in Fig. 2(b)). These calculations thus already provide an answer to the question of the topological class of 5-AGNRs and confirm the topological nature of their experimentally observed in-gap states. However, they do not clarify the mechanism for why the ribbons open a bandgap that relaxes towards the topological regime, since bond length relaxations on their own cannot be held responsible for it (purple point in Fig. 2(b)).

We propose that the bandgap opening can be driven by the different electrostatic potential felt by valence electrons at different regions of the ribbon due to the positive partial charge on the hydrogen atoms along the sides of the GNR. The importance of this effect has been tested by performing DFT calculations on 5-AGNRs with additional point charges centered in between the two hydrogen atoms of each unit cell on either GNR side (see inset in Fig. 2(c)). The point charges are modelled by a Yukawa potential with a decay constant $\lambda=0.01$ Å$^{-1}$ in order to avoid numerical instabilities due to the long range of the unscreened Coulomb interaction. Again, a qualitatively similar behaviour is observed as in the calculations shown in Fig. 2(a) and (b). A zero bandgap and topological inversion point is found for a charge of -1.04 |e| on each Yukawa potential center. More negative charges open a bandgap toward the trivial regime, and less negative charges open a bandgap toward the topological regime.

An analogy can again be drawn to the tight binding results of Fig. 2(a). A more negative charge will cause repulsion on the delocalized π-electrons and disfavour their localization at the bonds that are the closest to the Yukawa potential centers, namely the parallel bonds between the hydrogenated carbon atoms at the edge of the ribbon. This leads to an effectively lower $t_{par}$. Instead, a positive charge will attract the π-electrons towards those parallel bonds, increasing $t_{par}$ with respect to $t_{perp}$. At Q=0, the electrostatic potential generated by the positive partial charges of the hydrogen atoms already favours the π-electrons on those bonds and the inversion point at -1.04 |e| can be seen as the charge required to compensate for the effect of the hydrogen atoms. That is, this numerical experiment indicates that the ultimate reason for why the 5-aGNRs open a gap towards the topological side is the electrostatic potential generated by the partial charges of the hydrogen atoms along the GNR edges.

Overall, strain and the electrostatic potential along the GNR´s edges being potentially controllable parameters, this insight raises new possibilities to tune GNR´s topology and its associated properties.

In order to characterize the spatial variations of the electronic states of the 5-AGNRs, dI/dV point spectra were taken along one longitudinal side of a GNR and stacked into a colour-coded map. Such data are displayed in Fig. 3(a) for a 12-UC 5-AGNR adsorbed on Au(111). A number of discrete states are resolved, with their spatial distribution mapped out in a series of constant height dI/dV images that are presented in Fig. 3(b). Simulated constant-height STM images at 4 Å above the molecular plane (to account for the tip-sample distance and the locally different orbital decay into the vacuum[11] are shown in Fig. 3(c) (see methods) and the corresponding wavefunctions are displayed in Fig. 3(d). The excellent agreement between experiment and theory demonstrates that molecule-substrate coupling has little effect on the apparent local density of states on the nanoribbons, which allows for a straightforward assignment of the observed molecular states. They correspond to quantized states of the valence and conduction band, whose number of nodes increases with energy away from the Fermi level. It is important to note the difference in appearance between the onset of the valence and conduction bands in the dI/dV signal – in particular that, whereas the valence band onset is measured with highest intensity over the central region of the ribbon, as normally expected for the lowest energy quantized mode, the conduction band onset appears with a central node. This relates to the two different orbital symmetries that can be seen in the calculated gas phase orbitals (Fig. 3(d)). The onsets of the valence and conduction bands have this structure inverted, such that the valence band onset's central features are similar to those seen at the ends of the ribbon in the onset of the conduction band. Phase cancellation effects cause one of these types of orbital structure to extend much more into the vacuum than the other (especially along the sides of the ribbon). As a result, the STM signal, which probes the density of states several Å above the molecular plane, is greatly affected by the orbital's symmetry, as clearly supported by the simulated constant-height STM images shown in Fig. 3(c).

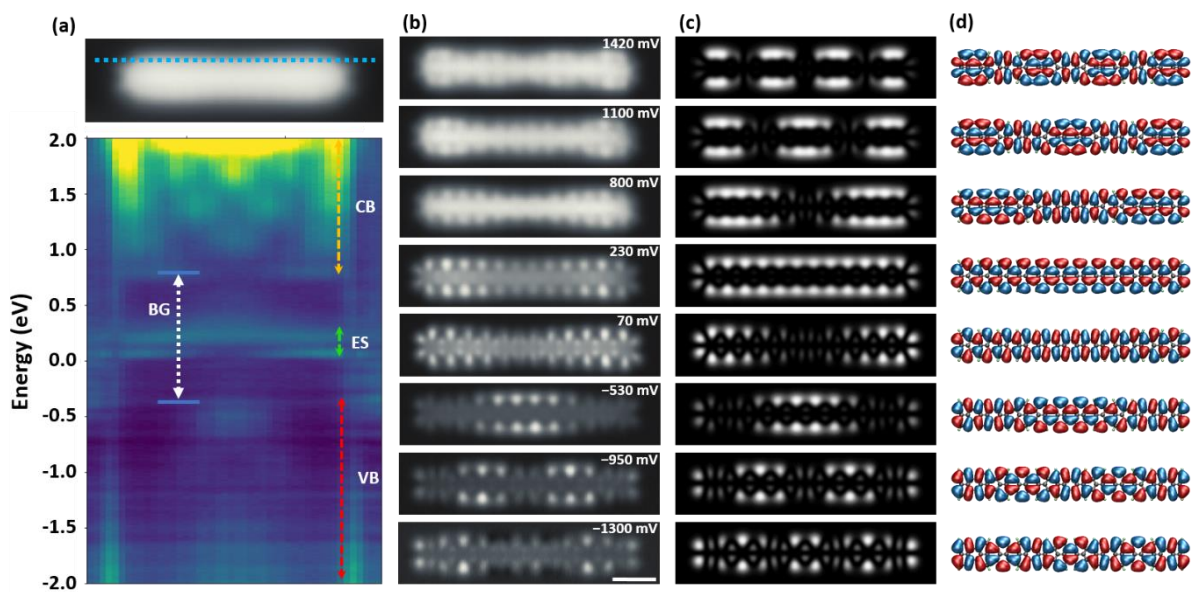

**Figure 3.** (a) Stacked dI/dV point spectra (40 points) along the side of a 12-UC 5-AGNR, as marked in the GNR STM image above, recorded with a metallic tip (Feedback opened at 303 pA, 500 mV. Oscillation: f = 731 Hz, A = 15 mV). The onsets of the valence and conduction bands (VB and CB) are indicated, as are the two in-gap 'end states'. (b) The corresponding constant height dI/dV images (Cl tip), with the associated bias voltage of each state. (c) Simulated STM images of the frontier states of a 12-UC 5-AGNR and (d) and the associated orbitals of free-standing GNRs.

Within the band gap of the 12-UC ribbon, two distinct states are clearly observed in the dI/dV spectra. These can be assigned to the end states of the nanoribbons, of which the lower energy state is occupied when the ribbon is in the gas phase. To examine the relationship between the length of the 5-AGNRs and the character of these in-gap states, we have measured a series of stacked dI/dV point spectra on ribbons of varying length. The data shown in Fig. 4 are for ribbons that have been manipulated with the STM tip and dragged onto 1 monolayer (ML) NaCl islands in a similar manner to experiments seen in the literature for 7-AGNRs[21,28], mixed 7/14-AGNRs[29], chiral GNRs[28] and zigzag GNRs[23] (see methods and SI). The insulating NaCl islands act as a decoupling layer that reduces the metal-molecule hybridization, which allows for the acquisition of clearer signals in dI/dV spectra and mapping[30] than the equivalent measurements on the metallic substrate. This in turn allows a more precise characterisation of the electronic structure of the ribbons, with better-defined peaks in the spectra. For comparison, dI/dV line scans and images of nanoribbons adsorbed on Au(111) are presented in Fig. S3, as well as in Fig. S4 with side-by-side 12-UC nanoribbon data on Au(111) and NaCl. More dI/dV images of the latter are also presented in Fig. S5.

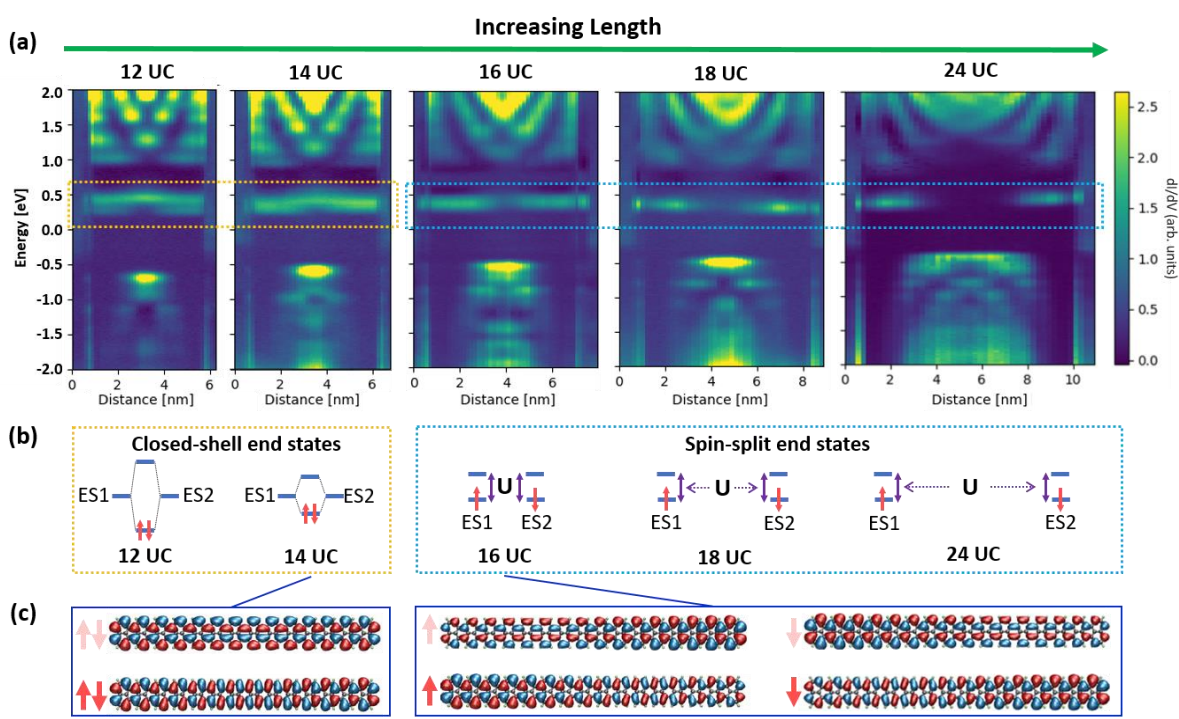

**Figure 4.** (a) A series of dI/dV line scans performed on 5-AGNRs of increasing length, all adsorbed on NaCl/Au(111)). The line scans are each scaled to represent the actual length of the ribbons. End states are highlighted with yellow and blue boxes. (b) A simple model for the occupation of the end states (ES1 and ES2) in free standing ribbons. In ribbons of 14-UC and below the two end states hybridize into bonding and antibonding orbitals, generating a closed-shell structure. In ribbons of 16-UC and above, the end states are degenerate singly occupied states, separated by the electron correlation energy U from their unoccupied counterparts. On Au and NaCl, all of the displayed end states are unoccupied due to a charge transfer to the substrate. (c) Gas phase molecular orbitals for 14-UC and 16-UC 5-AGNRs. Faded red arrows indicate unoccupied states in the gas phase.

A notable length-dependence is observed in the ribbon's electronic properties that only tends towards saturation for the longest ribbons that were measured (beyond 20 unit cells, Fig. S6). Of particular interest is the trend seen in the appearance of the end states of the ribbons as they change in length (dashed orange and blue boxes, Fig. 4(a)). On longer 5-AGNRs a single end state peak is observed that has most of its intensity at the termini of the ribbons. As the length decreases to 14-UC and below, this splits into two states with an energy gap that increases as the ribbons shorten. This threshold matches well with DFT calculations that predict a transition from degenerate singly occupied, spin-split end states to closed shell end states as they become closer and hybridize in the shorter ribbons. Simple schematics for this are presented in Fig. 4(b) and (c).

Focusing now on the longer open-shell ribbons, the Anderson model for magnetic states on metal substrates[31,32] predicts two resonances in the density of states: one below the Fermi level for the singly occupied state (SOMO), and another above the Fermi level (SUMO), separated from the occupied state by the electron correlation energy U. However, only one signal is observed in this case due to the high work function substrate, which causes the emptying of the end states (Fig. S6) by electron transfer from the GNR to the Au.

To further characterize 5-AGNRs and experimentally prove the magnetic nature of their topological end states, we performed transport measurements across GNRs in a two-terminal device structure. This is achieved using the STM tip and the substrate as the two contacts, with the GNR bridging the two.[22,33] This process is depicted in Fig. 5(a), in which the metallic STM tip is approached to one terminus of the nanoribbon with a low bias voltage (feedback open). As the tip approaches, the tunnelling barrier decreases and the associated current increases exponentially. A linear trace is thus obtained in a logarithmic plot of the current. Its slope β, also called the tunnelling decay constant, relates to the effective tunnelling barrier height and in a first approximation is associated with the average of the tip and sample work functions. After reaching a sufficiently close distance, a sudden jump in the current signifies the jump to contact of the ribbon to the tip. Thereafter the tip can be retracted, lifting the ribbon and resulting in the two-terminal device structure. As the ribbon is lifted, the current decreases with a dramatically reduced β constant (0.94±0.1 nm$^{-1}$ at 5 mV bias, Fig. 5(a)), which is among the lowest values reported for organic materials[34–36] and underlines the excellent conductive properties of these GNRs. Such a reduced β value can be understood from the low molecular band gap of 5-AGNRs, since the tunnelling barrier is now determined by the energy difference between $E_F$ and the frontier molecular orbitals, in combination with the renormalized, low effective mass of its charge carriers, which is an additional parameter that also determines β.[22]

Interestingly, at low lifting heights up to around 1 nm, an initial increase in the current is observed. While similar findings have been observed in previous experiments with 5-AGNRs[37], their explanation has so far remained unclear. To obtain a better understanding, we performed the lifting process in a stepwise manner and took dI/dV spectra at each step. Because of the high conductance displayed by these ribbons, these experiments were performed with ribbons atop a NaCl monolayer. The NaCl layer provides an additional tunnelling barrier, which reduces the current flow through the device structure and thus allows the probing of a larger bias window.

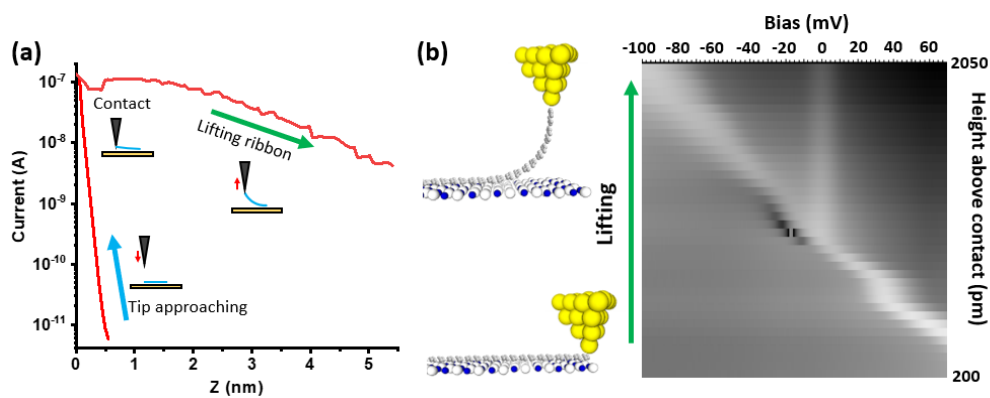

**Figure 5.** 5-AGNR lifting experiments. (a) Logarithmic plot of the current vs. relative tip height for the lifting of a 14-UC 5-AGNR on Au(111), showing the sudden change in conductance once the ribbon has been contacted with the STM tip. The zero-value of the relative tip height is set as the contact point. Schematics of the different lifting experiment scenarios are added as insets. (b) 2D conductance map of dI/dV spectra recorded during the lifting of a 22-UC 5-AGNR over NaCl. Each spectrum was recorded after moving to the next height value (steps of 50 pm). dI/dV intensity is plotted with a logarithmic scale to enhance the visibility of the zero-bias Kondo feature. Oscillation parameters: 731 Hz, 5 mV.

A series of dI/dV spectra for the lifting of a 22-UC nanoribbon is presented as a 2D conductance map in Fig. 5(b). With increasing lifting height, a peak in the dI/dV spectra observed at positive voltages shifts downwards in energy. This signal corresponds to the unoccupied end state, which for these 5-AGNRs ribbons extends several nm along the ribbon due to its low band gap, and can thus contribute to the charge transport at small lifting heights. As the end state approaches $E_F$ the tunnelling barrier is lowered and the current thus increases. Thereafter it crosses and moves away from $E_F$ again, causing the opposite effect. Ultimately the end state does not contribute to the charge transport as the lifting height surpasses its extension along the ribbon. This overall scenario explains the non-monotonic current dependence on the lifting height.

In addition, a new zero-bias peak is also observed alongside the now-occupied end state that continues to down-shift. It has a FWHM of approximately 14 meV, which, in the absence of a temperature-dependent analysis, is tenuously assigned to a Kondo temperature of around 82 K. To prove the Kondo nature of a zero bias peak, it needs to be shown that it follows the appropriate behaviour as a function of a particular parameter that distinguishes it from all other possible zero bias resonance sources. The most conventional method consists of sweeping the temperature and analysing its effect on the peak width, or sweeping a magnetic field and monitoring the expected Kondo resonance splitting. The former is unfeasible because of the uncontrollable thermal expansion in the "lifting geometry", while for the latter our Kondo peak is too broad to reveal conclusive splitting or broadening upon application of magnetic fields typically available in STMs. Instead, we sweep a parameter not generally available in STM experiments, namely the state occupancy. We sweep across the various occupation regimes described in the Anderson model for magnetic states[32], from the empty orbital regime at low lifting heights, to the mixed valence regime as the state crosses the Fermi level, finally reaching the Kondo regime as the state becomes populated with a full electron. The appearance of a zero bias resonance when the end state becomes occupied is unequivocally related with the Kondo effect, in contrast to other possible zero bias peak sources like tip artifacts. Indeed, similar occupation sweeps have been studied through gating in three-terminal devices[38,39], including gateable STM setups.[40] Here we mimic the gating by the fading electrostatic influence of the high work function substrate on the magnetic end state as it is lifted away from the surface.

The appearance of the Kondo peak upon the occupation of the end states is proof of their magnetism. However, the Kondo peak is also seen in lifting experiments with shorter ribbons that are not expected to have singly occupied end states. This initially unexpected finding actually relates to a change in end state to end state hybridisation when the ribbons are in a lifting geometry. The contact to the tip generates an asymmetry in the ribbon that changes the energies of the end states on either side, hindering their hybridization and leading to their open-shell character. This is indeed shown with a simple model in which the asymmetry is introduced by adding a hydrogen atom on one end of a 14-UC long ribbon. The DFT calculations on asymmetrically hydrogenated and pristine GNRs confirm the closed-shell character of the latter and the open-shell character of the former (Fig. S7). Furthermore it also confirms that the end state can survive in spite of an additional covalent bond to the carbon backbone (as is presumably present between the STM tip and GNR) and the associated $sp^3$ rehybridization of the associated carbon atom (Fig. S7 and Fig. S8).

What remains unclear from these experiments is which end state is being probed during the lifting: the state associated with the end of the ribbon in contact with the tip (through a bond of an unknown nature); or the end state of the ribbon terminus that is still adsorbed on the NaCl layer during lifting.

For the former, the Kondo screening would come from the metal tip electrons, while for the latter it would be from the Au substrate through the NaCl layer. To clarify this, and also to further support that these end states are singly occupied and thus reactive π-radicals, we exposed the ribbons at room temperature to molecular oxygen in the ultra-high vacuum chamber to a pressure of 2 x 10$^{-5}$ mbar for approximately 2 hours.

After the oxygen exposure, many ribbons had a significantly different appearance in the STM imaging, especially at positive bias voltages. Studying these ribbons via HR-STM (Fig. 6(a) and (b)) reveals a change in structure at the ends of the ribbon (a comparison is shown in Fig. S9), with what may be the addition of a new functional group. We tentatively suggest that this is a newly-formed ketone group, as depicted in Fig. 6(c). A stack of dI/dV spectra along a ribbon with this group at both ends (i.e. doubly oxidised) is shown in Fig. 6(d). In stark contrast to the pristine ribbon, no in-gap states are found in the doubly oxidised nanoribbon, suggesting that the oxidation reaction results in the destruction of the end states. Both valence and conduction bands are clearly visible, with small changes to their appearances. In particular, an increased intensity at the ends of the ribbon is observed throughout both bands.

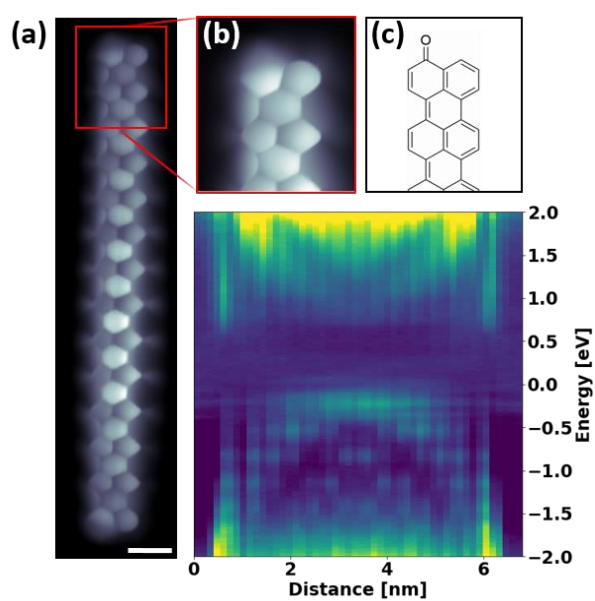

**Figure 6.** (a) HR-STM (5 mV, CO tip) of a doubly-oxidised 14-UC 5-AGNR (scale bar = 5 Å). (b) HR-STM zoom of the oxidised end of the same nanoribbon showing the change in contrast at the edge of one ring. (c) The suggested chemical structure for the oxidised NRs. (d) dI/dV line scan (oscillation parameters of 731 Hz, 15 mV) of the same doubly-oxidised nanoribbon. Both valence and conduction bands are visible, but there are no longer any in-gap end states.

Some of the nanoribbons were instead only half-oxidised after the exposure, presenting a perfect opportunity for investigating the source of the Kondo feature that is observed when lifting pristine 5-AGNRs. A half-oxidised 5-AGNR is presented in Fig. 7(a). It is clear from STM imaging at positive bias voltages that the pristine end still retains its end state, as is confirmed by the stacked dI/dV spectra, with the in-gap end state only detected on the pristine end of the ribbon. Notably, the centre of the valence and conduction bands is also spatially shifted relative to the geometrical centre of the nanoribbon, with the valence band presenting a higher intensity closer to the oxidised end, and vice-versa for the conduction band. This is supported by DFT calculations of the gas-phase molecular orbitals and simulated dI/dV images of a 16-UC half-oxidised 5-AGNR – these are presented alongside experimental constant height dI/dV images of the same nanoribbon in Fig. S10, with a high level of agreement between experiment and theory that further confirms the ketone group as the product of the oxidation.

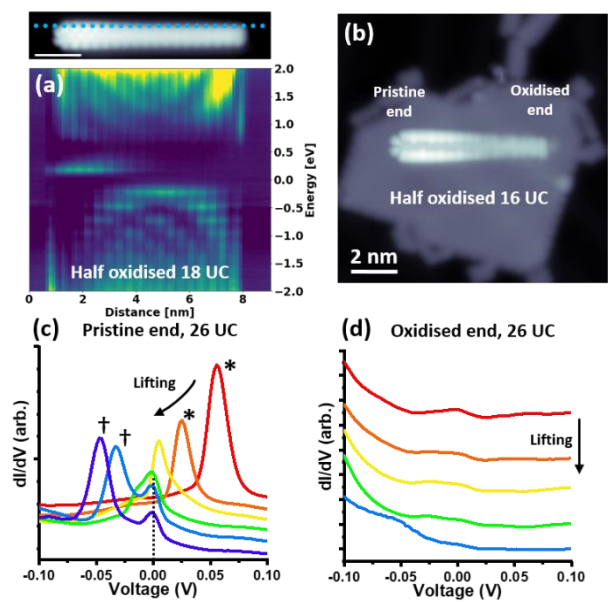

**Figure 7.** (a) A half-oxidised 18-UC 5-AGNR on Au(111), with a stacked dI/dV line scan plot recorded along one of its edges. The in-gap end state is only observed on the pristine end. Scale bar = 2 nm. (b) STM image (0.5 V) of a half-oxidised 16-UC 5-AGNR that has been dragged onto a 1 ML NaCl island via the STM tip, with the pristine and oxidised ends indicated. (c) and (d) dI/dV lifting experiments of another (26-UC) half-oxidised 5-AGNR on 1 ML NaCl, each via different ends of the ribbon. The end state shift and corresponding Kondo feature is only observed when lifting from the pristine end of the nanoribbon. Images of the nanoribbon used for these lifting experiments are presented in Fig. S11. † = 2x scale, * = 0.5x scale (to more clearly show the presence of the Kondo peak, which has a much lower intensity than the molecular state).

Half oxidized ribbons can also be manipulated onto a 1 ML NaCl island, as shown in Fig. 7(b). Lifting such nanoribbons provides clear evidence for the origin of the Kondo feature; it is only seen when lifting the pristine end of a half-oxidised ribbon, as shown in Fig. 7(c) and (d). This suggests that the origin of the observed end state that downshifts when lifting the pristine ends of 5-AGNRs is the end of the ribbon in contact with the tip. The increasing decoupling of the end state from the substrate as it is lifted by the tip shifts the state down towards its gas phase singly occupied state, with Kondo screening coming from the tip itself.

**Conclusions**

Although conventional tight binding calculations predict semi-metallic behavior for 5-AGNRs, through extensive experimental and theoretical studies we have demonstrated that 5-AGNRs on Au(111) display a semiconducting gap of 0.85 eV. The anisotropic electron potential generated by the partial charges of the hydrogen atoms along the sides of the 5-AGNRs favours the opening of a gap towards the topological (not the trivial) band structure. As a result, the nanoribbons have topological in-gap states that undergo a transition from singly occupied spin-split states to a closed-shell form below a critical GNR length of 14 unit cells and below. Although the end states are unoccupied when supported on Au(111) or on NaCl monolayers on Au(111), they become filled as the electrostatic influence of the high work function substrate fades when lifting the nanoribbons from the surface via STM tip manipulation. As the states become occupied a Kondo resonance appears, providing direct proof of their magnetic nature. The end states are no longer observed after GNR oxidation. Lifting experiments with half-oxidised ribbons demonstrate that the source of the observed Kondo feature is the end attached to the STM tip, as it is only observed when lifting a pristine ribbon end and not an oxidised one. We hope that these findings will further advance the understanding of the nature of such narrow AGNRs, as well as encourage others to consider and investigate the effects of topological states in other related systems. Precise understanding on the origin of topological states in such small-gap systems as 5-AGNRs may open the way towards adding a new functionality to graphene nanostructures: The results presented here indicate that these systems can be driven both towards their trivial and non-trivial phases using appropriate degrees of freedom and external parameters. In particular, our calculations here point towards strain and the electrostatic potential along the ribbon´s edges as two of them.

**Methods:**

STM:

All STM experiments were performed using a commercial Scienta-Omicron LT-STM, cooled to 4.3 K. A single crystal Au(111) sample (MaTeck) was used for all experiments. The crystal was cleaned via cycles of argon sputtering and annealing up to 720 K. The dibromoperylene (DBP) molecules were sublimed at 443 K onto the Au(111) crystal that was held at approximately 418 K during the deposition. NaCl was sublimed at 780 K on to the Au(111) crystal shortly after it was removed from the LT-STM. The exact sample temperature during this deposition is unknown, but it is believed to have been below room temperature, leading to the formation of 1 ML NaCl islands. Performing the same deposition at room temperature leads to bilayer islands. After NaCl deposition, the sample was placed inside the LT-STM. CO was then deposited onto the sample via a leak valve at a pressure of approximately $5 \times 10^{-9}$ mbar and a maximum sample temperature of 7.0 K. CO is usually imaged as a depression in the NaCl islands, and can be picked up with a metallic tip when scanning over it or by applying a -2 V bias voltage pulse when the tip is in typical tunnelling conditions (e.g. 100 pA, 0.5 V). Functionalisation of the tip with a Cl atom is achieved by approaching the tip by 350 pm from initial stabilisation conditions of 100 pA and 100 mV with the feedback off. After pickup, an increase in resolution is seen and a vacancy in the NaCl island can also be observed. dI/dV experiments were typically performed using a digital lock-in, with an oscillation frequency of 731 Hz and an amplitude of 5-20 mV.

Ribbon manipulation was typically achieved by approaching a metallic tip to the terminus of a ribbon with a low (5-10 mV) bias voltage. A sudden jump in current usually indicated the jump to contact, after which the ribbon was either directly lifted or dragged over the surface, followed by lifting. A sudden drop in the current usually indicates that the ribbon has fallen off the tip. To drag the ribbon onto 1 ML NaCl from the Au(111) surface, the ribbons were usually lifted at least 500 pm above the contact point before being pulled over the NaCl island. The absolute lifting height is unknown, as it is not clear how much of a jump the ribbon makes when forming a contact with the tip.

DFT:

First-principles electronic structure calculations were performed using Density Functional Theory (DFT) as implemented in SIESTA package.[41,42] Here we used the van der Waals density functional by Dion et al[43] with the modified exchange correlation by Klimeš, Bowler and Michaelides.[44] The valence electrons were described by a double-ζ plus polarization (DZP) basis set with the orbital radii defined using a 30 meV energy shift[42], while the core electrons were described using norm-conserving Troullier-Martins pseudopotentials.[45] For the calculations involving the metallic substrate, the Au(111) surface was represented by a 3-layer thick slab, where an extended DZP basis optimized for the description of the Au(111) surface were adopted for the top Au atomic layer.[46] Moreover, to obtain a reasonable energy position of the surface state in spite of the very this slab utilized, we employed a hydrogen passivation of the bottom layer of the slab.[47] An energy cutoff of 300 Ry was used for integrations in real space.[42] The smearing of the electronic occupations was defined by an electronic temperature of 300 K with a Fermi-Dirac distribution. The self-consistency cycle was stopped when variations on the elements of the density matrix were less than $10^{-6}$ ($10^{-4}$ on Au) and less than $10^{-5}$ eV ($10^{-4}$ eV on Au) for the Hamiltonian matrix elements.

Free-standing systems were calculated within a simulation cell where at least 30 Å of vacuum space was considered in order to avoid interaction with the replicas from neighbouring cells. For the calculations on the Au substrate the cell dimensions in the slab plane were determined by the experimental lattice parameter and such that the adsorbed finite ribbon was at least 10 Å away from neighbouring cells' replicas, while 20 Å of vacuum space was adopted in the direction perpendicular to the surface. In all cases, geometry optimizations were performed using the conjugate gradient method until all forces were lower than 20 meV/Å. Simulations on Au kept the two bottom layers fixed. For infinite ribbons we used a 100 k-point mesh along the GNR's periodic direction. For the calculations involving the Au substrate 2 k-points were adopted along the shorter slab plane dimension (i.e. perpendicular to the adsorbed ribbon backbone).

**Supporting Information**:
Comparison of molecular orbitals on finite and infinite GNRs displaying lattice spacings of 4.33 Å (relaxed lattice spacing for infinite 5-AGNRs) and 4.60 Å. Stacked dI/dV point spectra and dI/dV images of frontier molecular orbitals of differently long 5-AGNRs on Au(111). Comparison of stacked dI/dV point spectra and dI/dV images of frontier molecular orbitals for GNRs of similar length on Au(111) and NaCl/Au(111). Manipulation process and spectroscopic data of a 12-UC 5-AGNR on NaCl(Au(111). Length dependent molecular orbital energies for 5-AGNRs from experimental data on Au(111) and NaCl/Au(111), as well as from theory calculations for GNRs adsorbed on Au(111) and free-standing. Comparison of low-energy molecular orbitals for pristine 14-UC 5-AGNRs and asymmetric 14-UC 5-AGNs hydrogenated on one end. Comparative effect of asymmetric hydrogenation on one 5-AGNR end on its low-energy molecular orbitals depending on the location of the hydrogenated C atom. High resolution STM images of pristine and oxidized GNR ends. Experimental and calculated dI/dV images of molecular orbitals in singly oxidized GNRs. Topographic constant current images of the singly oxidized 5-AGNR used for the measurements displayed in Fig. 7c and Fig. 7d.

**Acknowledgements:**
We acknowledge funding from the European Union's Horizon 2020 programme (Grant Agreement Nos. 635919 and 863098 from ERC and FET Open projects, respectively), from the Spanish MINECO (Grant Nos. FIS2017-83780-P and MAT2016-78293-C6) and from the University of the Basque Country (Grant IT1246-19). D. G. O. thanks the Alexander von Humboldt Foundation for supporting his research stay at the MPI, and Klaus Kern for hosting him. We thank Peter Liljeroth for his generous donation of the molecular precursors used in this study.

**References:**

(1) Talirz, L.; Ruffieux, P.; Fasel, R. On-Surface Synthesis of Atomically Precise Graphene Nanoribbons. *Adv. Mater.* **2016**, 6222–6231. https://doi.org/10.1002/adma.201505738.

(2) Cai, J.; Ruffieux, P.; Jaafar, R.; Bieri, M.; Braun, T.; Blankenburg, S.; Muoth, M.; Seitsonen, A. P.; Saleh, M.; Feng, X.; et al. Atomically Precise Bottom-up Fabrication of Graphene

Nanoribbons. *Nature* **2010**, *466* (7305), 470–473. https://doi.org/10.1038/nature09211.

(3) Amsharov, K.; Berger, R.; Chen, S.; Chi, L.; Feng, X.; Fischer, F. R.; Grill, L.; Hecht, S.; Hu, A.; Koch, M.; et al. *From Polyphenylenes to Nanographenes and Graphene Nanoribbons*; Müllen, K., Feng, X., Eds.; Springer, 2017; Vol. 278. https://doi.org/10.1007/978-3-319-64170-6.

(4) Corso, M.; Carbonell-Sanromà, E.; de Oteyza, D. G. Bottom-Up Fabrication of Atomically Precise Graphene Nanoribbons. In *On-Surface Synthesis II*; de Oteyza, D. G., Rogero, C., Eds.; Springer, 2018; pp 113–152. https://doi.org/10.1007/978-3-319-75810-7.

(5) Merino-Díez, N.; Garcia-Lekue, A.; Carbonell-Sanromà, E.; Li, J.; Corso, M.; Colazzo, L.; Sedona, F.; Sánchez-Portal, D.; Pascual, J. I.; De Oteyza, D. G. Width-Dependent Band Gap in Armchair Graphene Nanoribbons Reveals Fermi Level Pinning on Au(111). *ACS Nano* **2017**, *11* (11), 11661–11668. https://doi.org/10.1021/acsnano.7b06765.

(6) Kharche, N.; Meunier, V. Width and Crystal Orientation Dependent Band Gap Renormalization in Substrate-Supported Graphene Nanoribbons. *J. Phys. Chem. Lett.* **2016**, *7* (8), 1526–1533. https://doi.org/10.1021/acs.jpclett.6b00422.

(7) Cao, T.; Zhao, F.; Louie, S. G. Topological Phases in Graphene Nanoribbons: Junction States, Spin Centers, and Quantum Spin Chains. *Phys. Rev. Lett.* **2017**, *119* (7), 076401. https://doi.org/10.1103/PhysRevLett.119.076401.

(8) Kimouche, A.; Ervasti, M. M.; Drost, R.; Halonen, S.; Harju, A.; Joensuu, P. M.; Sainio, J.; Liljeroth, P. Ultra-Narrow Metallic Armchair Graphene Nanoribbons. *Nat. Commun.* **2015**, *6*, 10177. https://doi.org/10.1038/ncomms10177.

(9) Zhang, H.; Lin, H.; Sun, K.; Chen, L.; Zagranyarski, Y.; Aghdassi, N.; Duhm, S.; Li, Q.; Zhong, D.; Li, Y.; et al. On-Surface Synthesis of Rylene-Type Graphene Nanoribbons. *J. Am. Chem. Soc.* **2015**, *137* (12), 4022–4025. https://doi.org/10.1021/ja511995r.

(10) Ruffieux, P.; Cai, J.; Plumb, N. C.; Patthey, L.; Prezzi, D.; Ferretti, A.; Molinari, E.; Feng, X.; Müllen, K.; Pignedoli, C. a.; et al. Electronic Structure of Atomically Precise Graphene Nanoribbons. *ACS Nano* **2012**, *6* (8), 6930–6935. https://doi.org/10.1021/nn3021376.

(11) Söde, H.; Talirz, L.; Gröning, O.; Pignedoli, C. A.; Berger, R.; Feng, X.; Müllen, K.; Fasel, R.; Ruffieux, P. Electronic Band Dispersion of Graphene Nanoribbons via Fourier-Transformed Scanning Tunneling Spectroscopy. *Phys. Rev. B - Condens. Matter Mater. Phys.* **2015**, *91* (4), 045429. https://doi.org/10.1103/PhysRevB.91.045429.

(12) Chen, Y. C.; De Oteyza, D. G.; Pedramrazi, Z.; Chen, C.; Fischer, F. R.; Crommie, M. F. Tuning the Band Gap of Graphene Nanoribbons Synthesized from Molecular Precursors. *ACS Nano* **2013**, *7* (7), 6123–6128. https://doi.org/10.1021/nn401948e.

(13) Sun, K.; Ji, P.; Zhang, J.; Wang, J.; Li, X.; Xu, X.; Zhang, H.; Chi, L. On-Surface Synthesis of 8- and 10-Armchair Graphene Nanoribbons. *Small* **2019**, *15*, 1804526. https://doi.org/10.1002/smll.201804526.

(14) Talirz, L.; Söde, H.; Dumslaff, T.; Wang, S.; Sanchez-Valencia, J. R.; Liu, J.; Shinde, P.; Pignedoli, C. A.; Liang, L.; Meunier, V.; et al. On-Surface Synthesis and Characterization of 9-Atom Wide Armchair Graphene Nanoribbons. *ACS Nano* **2017**, *11* (2), 1380–1388. https://doi.org/10.1021/acsnano.6b06405.

(15) Jänsch, D.; Ivanov, I.; Zagranyarski, Y.; Duznovic, I.; Baumgarten, M.; Turchinovich, D.; Li, C.; Bonn, M.; Müllen, K. Ultra-Narrow Low-Bandgap Graphene Nanoribbons from Bromoperylenes—Synthesis and Terahertz-Spectroscopy. *Chem. - A Eur. J.* **2017**, *23* (20), 4870–4875. https://doi.org/10.1002/chem.201605859.

(16) Yazyev, O. V. A Guide to the Design of Electronic Properties of Graphene Nanoribbons. *Acc. Chem. Res.* **2013**, *46* (10), 2319–2328. https://doi.org/10.1021/ar3001487.

(17) Barone, V.; Hod, O.; Scuseria, G. E. Electronic Structure and Stability of Semiconducting Graphene Nanoribbons. *Nano Lett.* **2006**, *6* (12), 2748–2754. https://doi.org/10.1021/nl0617033.

(18) Son, Y.-W.; Cohen, M. L.; Louie, S. G. Energy Gaps in Graphene Nanoribbons. *Phys. Rev. Lett.* **2006**, *97* (21), 216803. https://doi.org/10.1103/PhysRevLett.97.216803.

(19) Rizzo, D. J.; Veber, G.; Cao, T.; Bronner, C.; Chen, T.; Zhao, F.; Rodriguez, H.; Louie, S. G.; Crommie, M. F.; Fischer, F. R. Topological Band Engineering of Graphene Nanoribbons. *Nature* **2018**, *560* (7717), 204–208. https://doi.org/10.1038/s41586-018-0376-8.

(20) Gröning, O.; Wang, S.; Yao, X.; Pignedoli, C. A.; Borin Barin, G.; Daniels, C.; Cupo, A.; Meunier, V.; Feng, X.; Narita, A.; et al. Engineering of Robust Topological Quantum Phases in Graphene Nanoribbons. *Nature* **2018**, *560* (7717), 209–213. https://doi.org/10.1038/s41586-018-0375-9.

(21) Wang, S.; Talirz, L.; Pignedoli, C. A.; Feng, X.; Müllen, K.; Fasel, R.; Ruffieux, P. Giant Edge State Splitting at Atomically Precise Graphene Zigzag Edges. *Nat. Commun.* **2016**, *7*, 11507. https://doi.org/10.1038/ncomms11507.

(22) Koch, M.; Ample, F.; Joachim, C.; Grill, L. Voltage-Dependent Conductance of a Single Graphene Nanoribbon. *Nat. Nanotechnol.* **2012**, *7*, 713–717. https://doi.org/10.1038/nnano.2012.169.

(23) Ruffieux, P.; Wang, S.; Yang, B.; Sanchez-Sanchez, C.; Liu, J.; Dienel, T.; Talirz, L.; Shinde, P.; Pignedoli, C. A.; Passerone, D.; et al. On-Surface Synthesis of Graphene Nanoribbons with Zigzag Edge Topology. *Nature* **2016**, *531* (7595), 489–492. https://doi.org/10.1038/nature17151.

(24) Wakabayashi, K.; Sasaki, K. I.; Nakanishi, T.; Enoki, T. Electronic States of Graphene Nanoribbons and Analytical Solutions. *Sci. Technol. Adv. Mater.* **2010**, *11* (5), 054504. https://doi.org/10.1088/1468-6996/11/5/054504.

(25) Lee, H.; Son, Y. W.; Park, N.; Han, S.; Yu, J. Magnetic Ordering at the Edges of Graphitic Fragments: Magnetic Tail Interactions between the Edge-Localized States. *Phys. Rev. B - Condens. Matter Mater. Phys.* **2005**, *72* (17), 174431. https://doi.org/10.1103/PhysRevB.72.174431.

(26) Jelinek, P. High Resolution SPM Imaging of Organic Molecules with Functionalized Tips. *J. Phys. Condens. Matter* **2017**, *29* (34), 343002. https://doi.org/10.1088/1361-648X/aa76c7.

(27) Gross, L.; Moll, N.; Mohn, F.; Curioni, A.; Meyer, G.; Hanke, F.; Persson, M. High-Resolution Molecular Orbital Imaging Using a p-Wave STM Tip. *Phys. Rev. Lett.* **2011**, *107* (8), 086101. https://doi.org/10.1103/PhysRevLett.107.086101.

(28) Jacobse, P. H.; Mangnus, M. J. J.; Zevenhuizen, S. J. M.; Swart, I. Mapping the Conductance of Electronically Decoupled Graphene Nanoribbons. *ACS Nano* **2018**, *12* (7), 7048–7056. https://doi.org/10.1021/acsnano.8b02770.

(29) Wang, S.; Kharche, N.; Costa Girão, E.; Feng, X.; Müllen, K.; Meunier, V.; Fasel, R.; Ruffieux, P. Quantum Dots in Graphene Nanoribbons. *Nano Lett.* **2017**, *17* (7), 4277–4283. https://doi.org/10.1021/acs.nanolett.7b01244.

(30) Repp, J.; Meyer, G.; Stojković, S. M.; Gourdon, A.; Joachim, C. Molecules on Insulating Films:

Scanning-Tunneling Microscopy Imaging of Individual Molecular Orbitals. *Phys. Rev. Lett.* **2005**, *94* (2), 026803. https://doi.org/10.1103/PhysRevLett.94.026803.

(31) Ternes, M.; Heinrich, A. J.; Schneider, W. D. Spectroscopic Manifestations of the Kondo Effect on Single Adatoms. *J. Phys. Condens. Matter* **2009**, *21* (5), 053001. https://doi.org/10.1088/0953-8984/21/5/053001.

(32) Anderson, P. W. Localized Magnetic States in Metals. *Phys. Rev.* **1961**, *124* (1), 41–53. https://doi.org/10.1063/1.1708389.

(33) Lafferentz, L.; Ample, F.; Yu, H.; Hecht, S.; Joachim, C.; Grill, L. Conductance of a Single Conjugated Polymer as a Continuous Function of Its Length. *Science* **2009**, *323* (5918), 1193–1197. https://doi.org/10.1126/science.1168738.

(34) Xiang, D.; Wang, X.; Jia, C.; Lee, T.; Guo, X. Molecular-Scale Electronics: From Concept to Function. *Chem. Rev.* **2016**, *116* (7), 4318–4440. https://doi.org/10.1021/acs.chemrev.5b00680.

(35) Su, T. A.; Neupane, M.; Steigerwald, M. L.; Venkataraman, L.; Nuckolls, C. Chemical Principles of Single-Molecule Electronics. *Nat. Rev. Mater.* **2016**, *1*, 16002. https://doi.org/10.1038/natrevmats.2016.2.

(36) Nacci, C.; Ample, F.; Bleger, D.; Hecht, S.; Joachim, C.; Grill, L. Conductance of a Single Flexible Molecular Wire Composed of Alternating Donor and Acceptor Units. *Nat. Commun.* **2015**, *6*, 7397. https://doi.org/10.1038/ncomms8397.

(37) Jacobse, P. H.; Kimouche, A.; Gebraad, T.; Ervasti, M. M.; Thijssen, J. M.; Liljeroth, P.; Swart, I. Electronic Components Embedded in a Single Graphene Nanoribbon. *Nat. Commun.* **2017**, *8*, 119. https://doi.org/10.1038/s41467-017-00195-2.

(38) Goldhaber-Gordon, D.; Göres, J.; Kastner, M. A.; Shtrikman, H.; Mahalu, D.; Meirav, U. From the Kondo Regime to the Mixed-Valence Regime in a Single-Electron Transistor. *Phys. Rev. Lett.* **1998**, *81* (23), 5225–5228. https://doi.org/10.1103/PhysRevLett.81.5225.

(39) Scott, G. D.; Hu, T. C. Gate-Controlled Kondo Effect in a Single-Molecule Transistor with Elliptical Ferromagnetic Leads. *Phys. Rev. B* **2017**, *96*, 144416. https://doi.org/10.1103/PhysRevB.96.144416.

(40) Jiang, Y.; Lo, P. W.; May, D.; Li, G.; Guo, G. Y.; Anders, F. B.; Taniguchi, T.; Watanabe, K.; Mao, J.; Andrei, E. Y. Inducing Kondo Screening of Vacancy Magnetic Moments in Graphene with Gating and Local Curvature. *Nat. Commun.* **2018**, *9*, 2349. https://doi.org/10.1038/s41467-018-04812-6.

(41) Artacho, E.; Sánchez-Portal, D.; Ordejón, P.; García, A.; Soler, J. M. Linear-Scaling Ab-Initio Calculations for Large and Complex Systems. *Phys. Status Solidi* **1999**, *215* (1), 809–817. https://doi.org/10.1002/(SICI)1521-3951(199909)215:1<809::AID-PSSB809>3.0.CO;2-0.

(42) Soler, J. M.; Artacho, E.; Gale, J. D.; García, A.; Junquera, J.; Ordejón, P.; Sánchez-Portal, D. The SIESTA Method for Ab Initio Order-N Materials Simulation. *J. Phys. Condens. Matter* **2002**, *14* (11), 2745–2779. https://doi.org/10.1088/0953-8984/14/11/302.

(43) Dion, M.; Rydberg, H.; Schröder, E.; Langreth, D. C.; Lundqvist, B. I. Van Der Waals Density Functional for General Geometries. *Phys. Rev. Lett.* **2004**, *92* (24), 246401. https://doi.org/10.1103/PhysRevLett.92.246401.

(44) Klimeš, J.; Bowler, D. R.; Michaelides, A. Chemical Accuracy for the van Der Waals Density Functional. *J. Phys. Condens. Matter* **2010**, *22* (2), 022201. https://doi.org/10.1088/0953-

8984/22/2/022201.

(45) Troullier, N.; Martins, J. L. Efficient Pseudopotentials for Plane-Wave Calculations. *Phys. Rev. B* **1991**, *43* (3), 1993–2006. https://doi.org/10.1103/PhysRevB.43.1993.

(46) García-Gil, S.; García, A.; Lorente, N.; Ordejón, P. Optimal Strictly Localized Basis Sets for Noble Metal Surfaces. *Phys. Rev. B - Condens. Matter Mater. Phys.* **2009**, *79* (7), 075441. https://doi.org/10.1103/PhysRevB.79.075441.

(47) Gonzalez-Lakunza, N.; Fernández-Torrente, I.; Franke, K. J.; Lorente, N.; Arnau, A.; Pascual, J. I. Formation of Dispersive Hybrid Bands at an Organic-Metal Interface. *Phys. Rev. Lett.* **2008**, *100*, 156805. https://doi.org/10.1103/PhysRevLett.100.156805.

# SUPPLEMENTARY INFORMATION

**Probing the magnetism of topological end-states in 5-armchair graphene nanoribbons**

James Lawrence,[1,2,†] Pedro Brandimarte,[1,†] Alejandro Berdonces,[1,2] Mohammed S. G. Mohammed,[1,2] Abhishek Grewal,[3] Christopher C. Leon,[3] Daniel Sanchez-Portal,[1,2,*] Dimas G. de Oteyza.[1,2,4,*]

[1] Donostia International Physics Center, 20018 San Sebastián, Spain

[2] Centro de Fisica de Materiales, CSIC-UPV/EHU, 20018 San Sebastián, Spain

[3] Max-Planck-Institut für Festkörperforschung, 70569 Stuttgart, Germany

[4] Ikerbasque, Basque Foundation for Science, 48011 Bilbao, Spain

[†] These authors contributed equally

[*] daniel.sanchez@ehu.eus, d_g_oteyza@ehu.eus

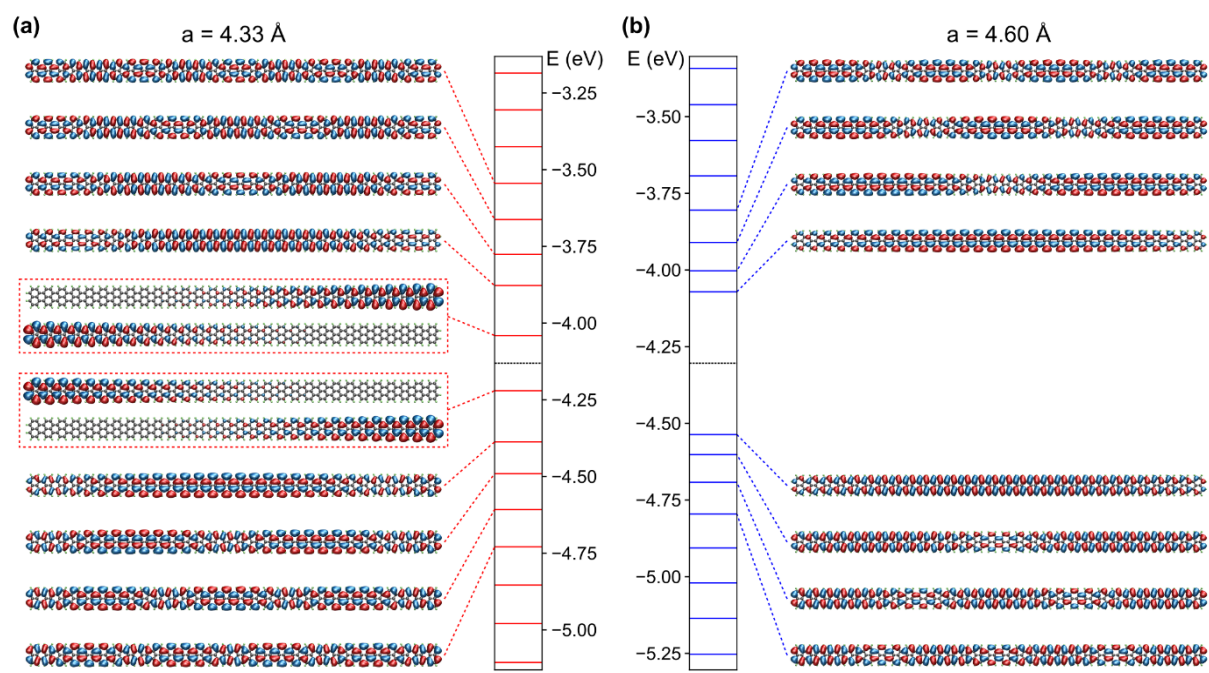

**Figure S1.** A comparison of gas-phase DFT calculations for the topologically (a) non-trivial and (b) trivial 30-UC 5-AGNR. The structures were built by taking the unit cell geometry optimized at a fixed lattice parameter (a = 4.33 Å for the non-trivial case and a = 4.60 Å for the trivial one, see Fig. S2), replicating it and only leaving it to further relax the hydrogen atoms. No spin-split end states can be seen on the trivial ribbon, and the orbital character of the valence and conduction band onsets are noticeably inverted from those seen in the relaxed ribbon (Fig. S2(a) and main paper Fig. 1(d)).

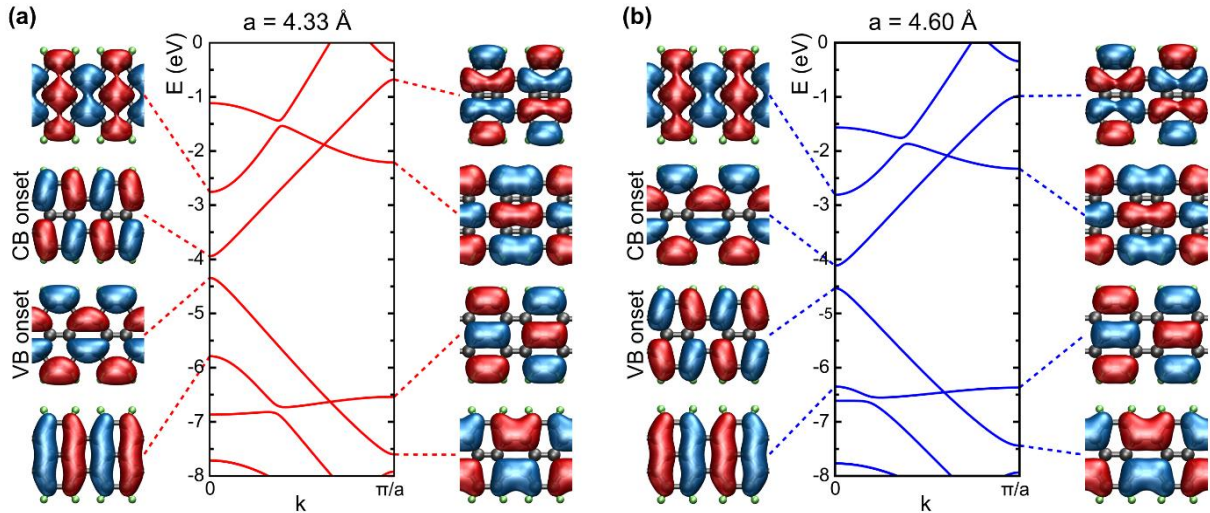

**Figure S2.** The calculated band structures of free-standing infinite relaxed 5-AGNRs. (a) Topologically non-trivial lowest energy configuration with lattice parameter a = 4.33 Å. (b) A topologically trivial configuration optimized at a fixed larger lattice parameter of a = 4.60 Å. The VB and CB onsets at the Γ point are exchanged when compared to the non-trivial case (a). The energies are given with respect to the vacuum level.

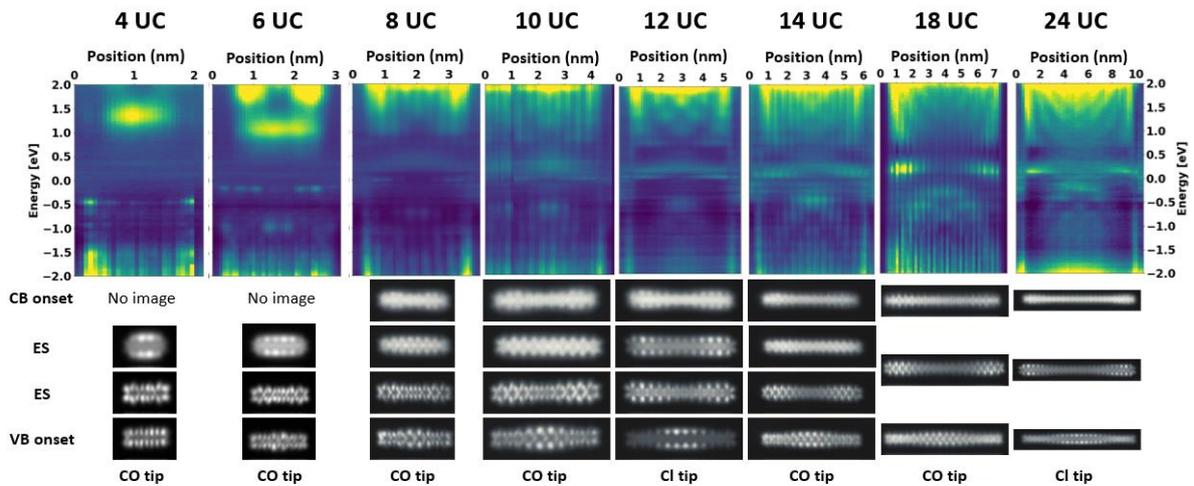

**Figure S3.** A series of dI/dV line scans for ribbons of increasing length, adsorbed on Au(111), accompanied by constant height dI/dV images of their CB onset, end states and VB onset (where possible). The tip used for the images is indicated below. The sharp transition seen in the 10-UC line scan is due to the ribbon moving part way through the experiment. However, all states of interest can still be seen in the region recorded before it shifted.

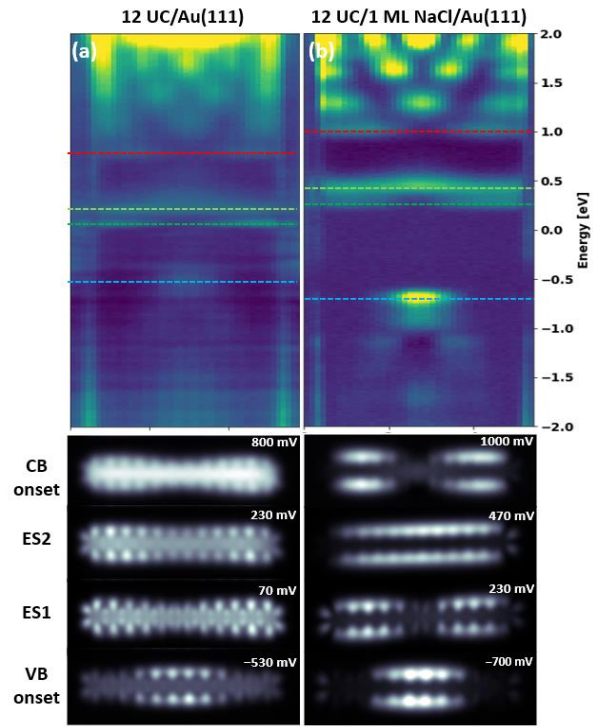

**Figure S4.** A comparison between dI/dV line scans and images of 12-UC ribbons on both Au(111) and 1ML NaCl islands. The states imaged in the lower half of the figure are indicated with dashed lines in the line scan. There is a clear increase in the band-gap of the ribbons when adsorbed on NaCl, as would be expected due to a reduced level of screening from the metallic substrate. However, there is also an upshift of both end-states relative to the centre of the band gap. In both cases the end states are unoccupied, that is, the ribbon is positively charged when it is adsorbed on Au(111) with or without the NaCl buffer layer.

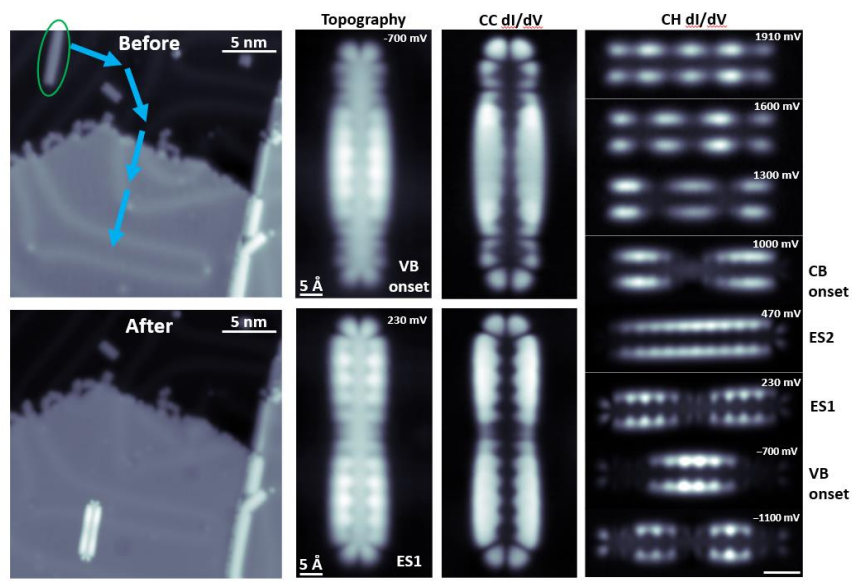

**Figure S5.** The manipulation and imaging of a 12-UC 5-AGNR onto a 1ML NaCl island. After making contact with the end of the ribbon with the STM tip, it was dragged along the path indicated by the blue arrows, and then lifted/dragged onto the NaCl island. Various dI/dV images of the same ribbon

are shown to the right, with both constant current (CC) and constant height (CH) images shown. The confinement effects on the electronic states of the conduction band can be much more clearly seen here, with an increasing number of nodes observed at higher bias voltages. The scale bar of the images on the right-hand side is 1 nm, and the images were recorded with a Cl-functionalized tip.

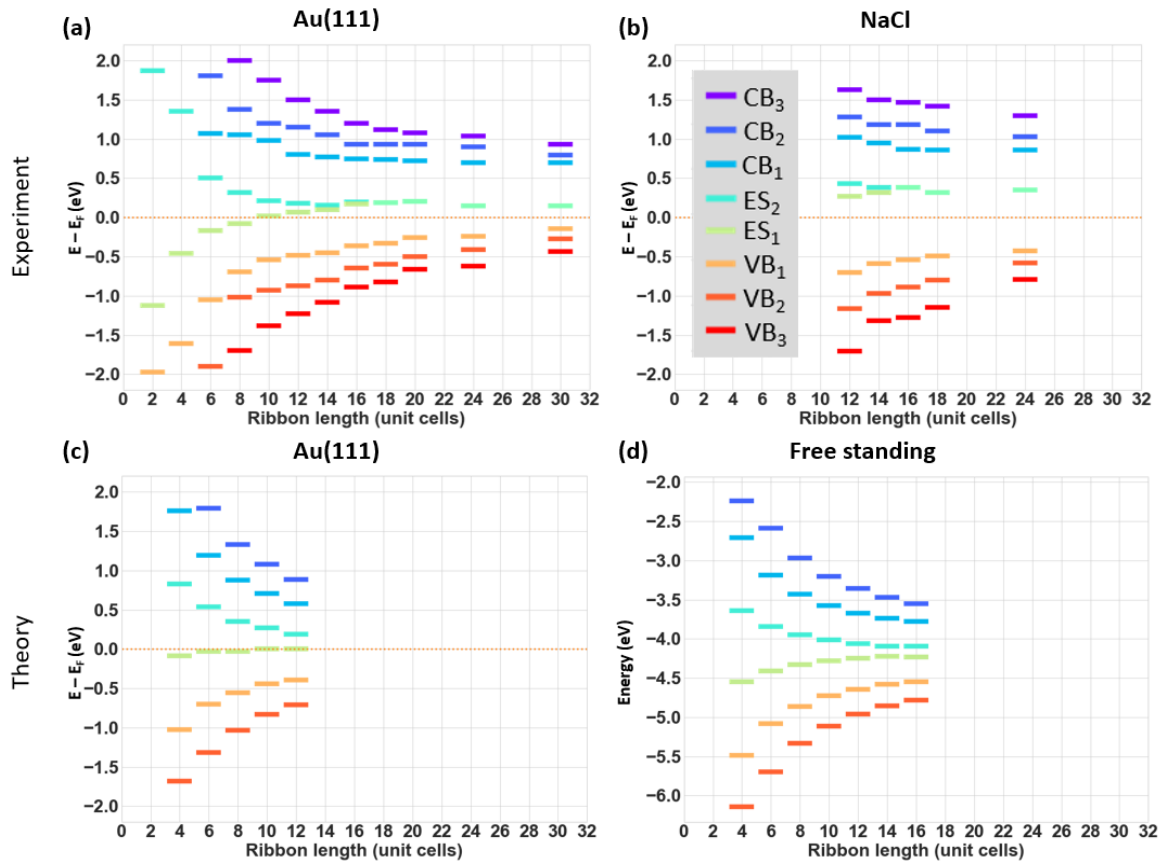

**Figure S6.** Energies for the lowest energy quantization states related to valence and conduction band, as well as for the in-gap end states as a function of the GNR length on Au(111) (a) and on a NaCl monolayer (b) (referred to $E_F$, marked with an orange dashed line at E=0). The energies of those same states from calculations on Au(111)-supported and free-standing GNRs are displayed in (c) and (d), respectively (referred to the Fermi level marked by an orange dashed line in (c) and with respect to the vacuum energy in (d)). Comparison of the data on Au(111) and NaCl/Au(111) clearly show the effect of the reduced screening (a larger apparent bandgap) on NaCl for all lengths. Beyond the different band gap values on either substrate, both datasets display a similar strong length dependence that only tends to saturation for the longest measured ribbons (beyond 20 unit cells). A comparable length dependence is observed in our calculations for both Au(111)-supported and free standing GNRs. Notice, however, that these DFT calculations do not account for all effects associated with the different screening by the different substrates, such as the energy renormalization due to the long range image potential. In addition, both on Au(111) and NaCl/Au(111) the GNRs get charged by electron donation to the high workfunction substrate for ribbons above a certain length threshold (8-10 UC on Au(111)). This is again confirmed by DFT calculations, which show a similar length threshold (10 UC) at which the GNRs become charged on Au(111). The hybridization with the Au(111) substrate provides the ribbon's levels with a finite width and in some cases makes their identification harder. However, a detailed comparison of the projected density of states on the ribbon with the energy

spectra from free-standing ribbons allows for an unambiguous identification of the levels in the DFT calculations.

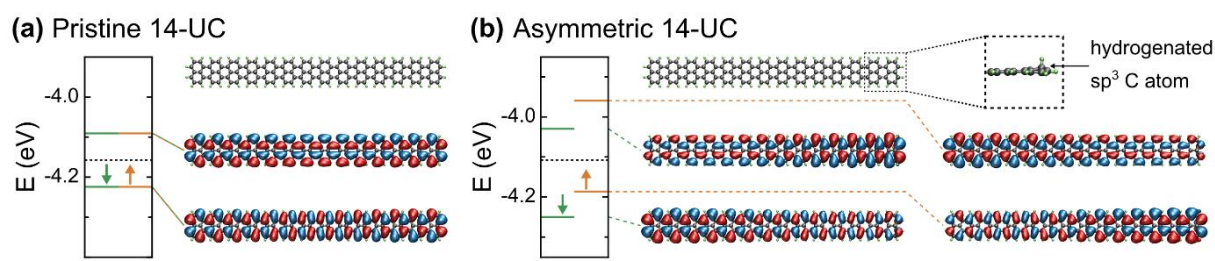

**Figure S7.** (a) Structure of a pristine 14-UC 5-AGNR and its associated closed-shell end state orbitals. (b) Structure of an asymmetric 14-UC 5-GNR with a hydrogenated C atom on one end and its associated open-shell end state orbitals. Orange and green lines indicate opposite spin.

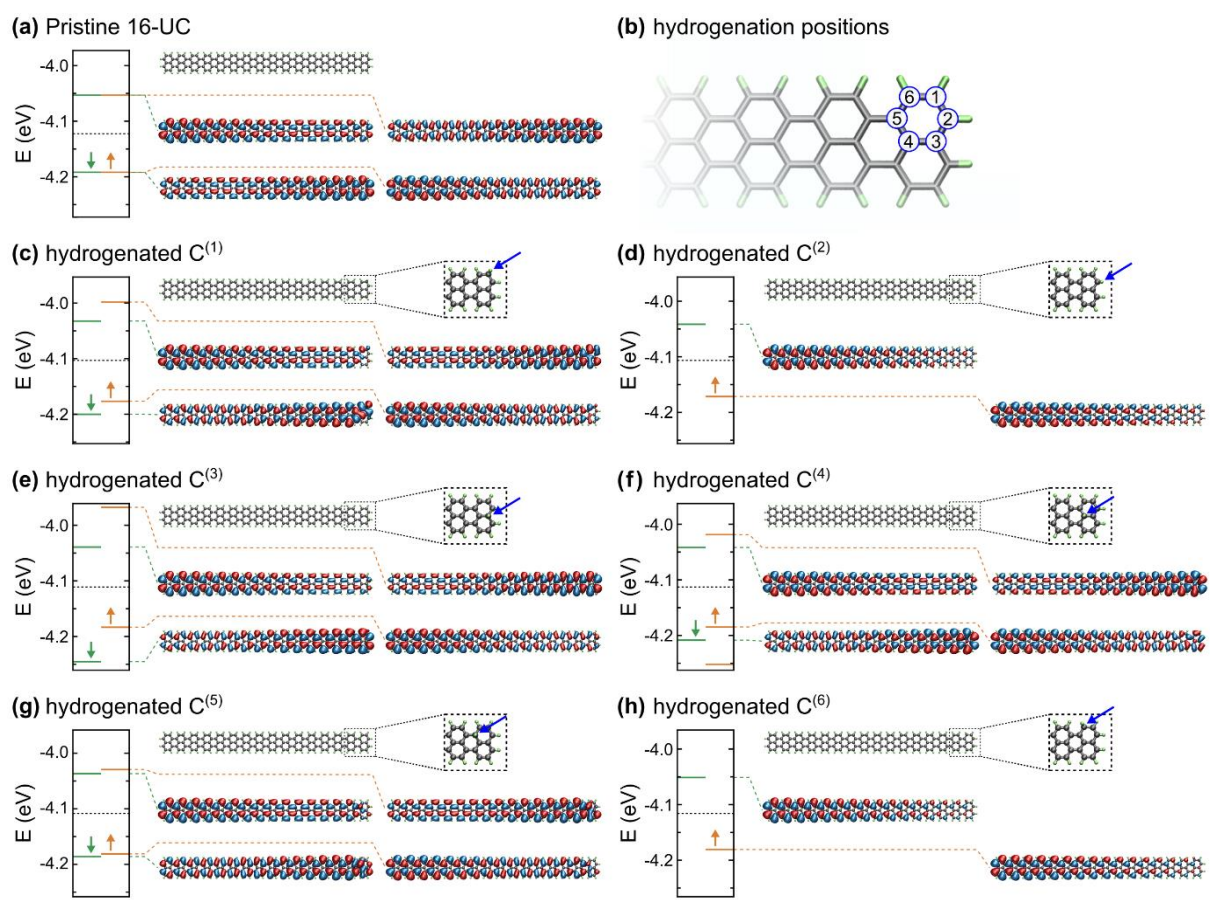

**Figure S8.** Effect of C atoms hydrogenation on the end states of a 16-UC 5-AGNR. (a) Structure of a pristine 16-UC 5-AGNR and its associated end state orbitals. (b) Different positions of the C atom considered for hydrogenation. (c-h) Structure and associated end state orbitals of a 16-UC 5-GNR with a hydrogenated C atom for each of the positions specified in panel (b). Of all 6 cases considered, only two (labeled as '2' and '6' on panel (b)) resulted in quenching the disturbed end state.

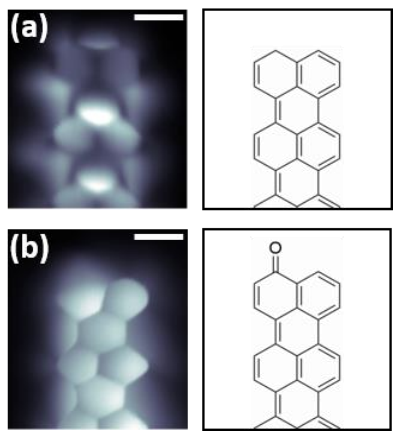

**Figure S9.** A comparison between constant height HR-STM images of (a) the pristine end of a 5-AGNR (CO tip, 2 mV) and (b) the oxidised end of a 5-AGNR (CO tip, 2 mV). Scale bars are both 3 Å. Due to the presence of the end state at low positive bias voltages, the appearance of the pristine nanoribbons (when above 8 UC in length) was typically a mixture of structural features and the end state. As oxidation destroys the end state, the appearance of oxidised ends is unaffected by this, and the structure is imaged much more clearly.

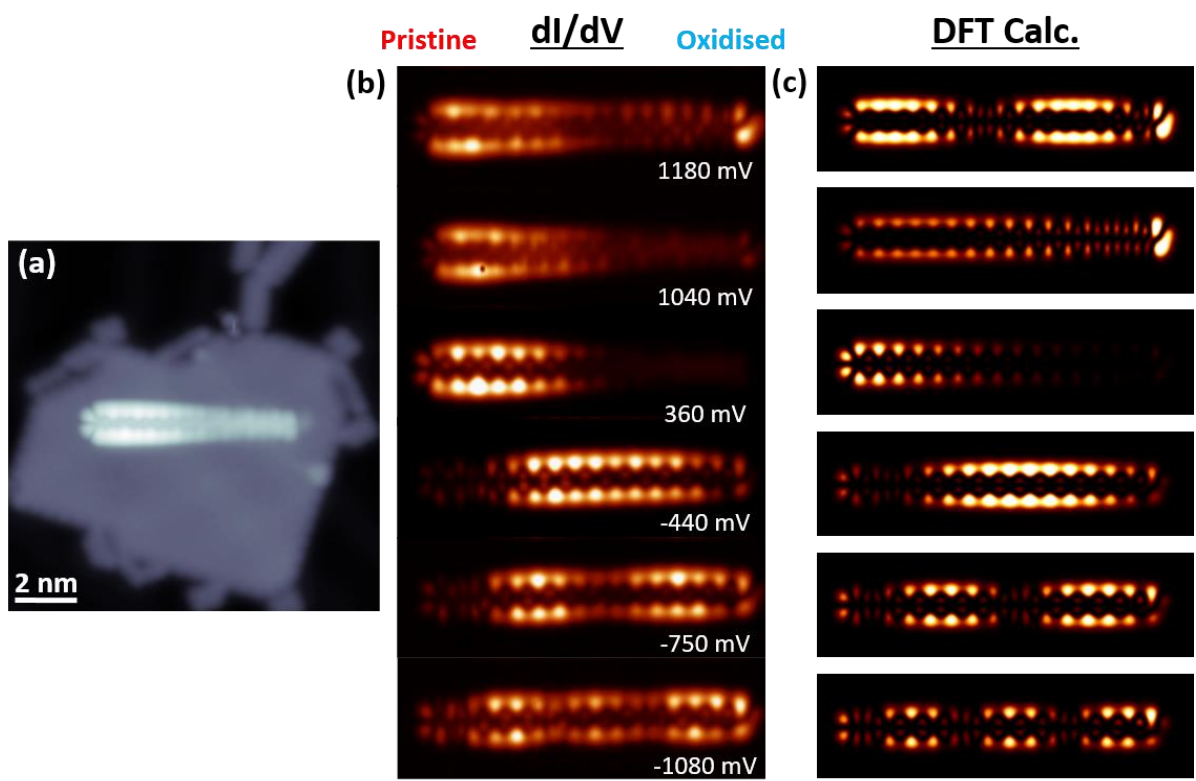

**Figure S10.** Constant height dI/dV images of the half-oxidised 16-UC nanoribbon shown in the main paper, with the pristine and oxidised ends indicated. In (c), DFT-calculated gas phase molecular orbitals of a half-oxidised 16-UC ribbon are shown. A generally good agreement is found between the images and the calculations. In particular, the (spatial) shift of the valence band towards the oxidised end is clear in both the experimental results and the calculations, with the opposite effect (i.e. a shift towards the pristine end) seen for the conduction band.

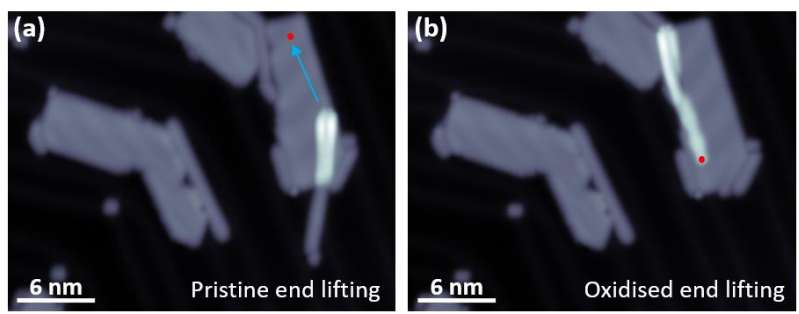

**Figure S11.** The lifting positions used for the half-oxidised 26-UC data presented in main paper Fig. 7 (c) and (d). (a) The ribbon was first dragged partially onto the NaCl island by its pristine end, and then dragged further in and lifted at the position indicated by its red marker. After this, it was released by a +2V pulse (using the methodology of Wang et al.[1]) and landed in the position shown in (b), in which it is decoupled by both the NaCl and also an adjacent nanoribbon. The ribbon was then lifted by its oxidised end, indicated again with a red marker. The exact position of the pristine end of the ribbon during this process is unknown, but it is assumed to be decoupled from the Au substrate due to the magnitude of the current exhibited. Typically when lifting directly from Au(111), the current was found to be of the order of $10^{-7}$ A when first making contact at 5 mV bias voltage, whereas when lifting on 1 ML NaCl it was generally in the $10^{-9}$ A range.

**Supplementary Methods:**

**Precursor Synthesis:**

The synthesis of the precursors used in this study is described in the supplementary information of the article by Kimouche et al.[2], who kindly also allowed us to use their precursors.

**References:**

(1) Wang, S.; Talirz, L.; Pignedoli, C. A.; Feng, X.; Müllen, K.; Fasel, R.; Ruffieux, P. Giant Edge State Splitting at Atomically Precise Graphene Zigzag Edges. *Nat. Commun.* **2016**, *7*, 11507. https://doi.org/10.1038/ncomms11507.

(2) Kimouche, A.; Ervasti, M. M.; Drost, R.; Halonen, S.; Harju, A.; Joensuu, P. M.; Sainio, J.; Liljeroth, P. Ultra-Narrow Metallic Armchair Graphene Nanoribbons. *Nat. Commun.* **2015**, *6*, 10177. https://doi.org/10.1038/ncomms10177.